\documentclass[twocolumn,superscriptaddress,showpacs,prl,aps,amsmath,amssymb,nofootinbib]{revtex4-1}                                        
\usepackage{natbib}
\usepackage{graphicx,color}
\usepackage{amsmath,amssymb}
\usepackage{verbatim}
\usepackage{float}
\usepackage{wasysym}
\usepackage{amssymb,graphicx}
\usepackage{epsfig}
\usepackage{psfrag}
\usepackage{dsfont}
\usepackage{amsfonts}
\usepackage{mathrsfs}
\usepackage{multirow}
\usepackage{times}
\usepackage{bm}
\usepackage{hyperref}
\usepackage{xspace}
\hypersetup{
  colorlinks=true,        
  linkcolor=blue,         
  citecolor=cyan,         
}
\usepackage{pifont}
\usepackage[normalem]{ulem}   

\graphicspath{{./}}

\newcommand{\beq}{\begin{equation}} 
\newcommand{\eeq}{\end{equation}} 
\newcommand{\beqn}{\begin{eqnarray}} 
\newcommand{\eeqn}{\end{eqnarray}}

\newcommand{\zD}{{\raise1.0ex\hbox{${}^{\ \circ}$}}\!\!\!\!\!D}
\newcommand{\alone}{{\raise0.5ex\hbox{${}^{\ 1}$}}\!\!\!\!\alpha}

\newcommand{\nalam}{\mathrel{\raise0.9ex\hbox{$^\lambda$}\mkern-14mu
\lower0.0ex\hbox{$\nabla$}}}

\newcommand{\zeroD}{{\raise1.0ex\hbox{${}^{\ \circ}$}}\!\!\!\!\!D}

\newcommand{\zLap}{{\raise1.0ex\hbox{${}^{\ \circ}$}}\!\!\!\!\Delta}
\newcommand{\zna}{{\raise1.0ex\hbox{${}^{\ \circ}$}}\!\!\!\!\!\nabla}
\newcommand{\zS}{{\raise1.0ex\hbox{${}^{\ \circ}$}}\!\!\!\!\!S}

\newcommand{\illinois}{\textsc{Illinois GRMHD}\xspace}

\newcommand{\GE}{\epsilon}

\newcommand{\GR}{\rho}
\newcommand{\GS}{\sigma}

\newcommand{\GP}{\phi}

\newcommand{\Ms}{M_\odot}
\newcommand{\cmark}{\ding{51}}%
\newcommand{\xmark}{\ding{55}}%

\newcommand{\be}{\begin{equation}}
\newcommand{\ee}{\end{equation}}

\def\QEQ{{%
    \setbox0\hbox{$I$}%
    \rlap{\hbox to \wd0{\hss--\hss}}\box0
}}

\begin{document}

\title{ Dynamically stable ergostars exist: General relativistic models and simulations  }

\author{Antonios Tsokaros}
\affiliation{Department of Physics, University of Illinois at Urbana-Champaign, Urbana, IL 61801}
\email{tsokaros@illinois.edu}

\author{Milton Ruiz}
\affiliation{Department of Physics, University of Illinois at Urbana-Champaign, Urbana, IL 61801}
\author{Lunan Sun}
\affiliation{Department of Physics, University of Illinois at Urbana-Champaign, Urbana, IL 61801}
\author{Stuart L. Shapiro}
\affiliation{Department of Physics, University of Illinois at Urbana-Champaign, Urbana, IL 61801}
\affiliation{Department of Astronomy \& NCSA, University of Illinois at Urbana-Champaign, Urbana, IL 61801}
\author{K\=oji Ury\=u}
\affiliation{Department of Physics, University of the Ryukyus, Senbaru, Nishihara, Okinawa 903-0213, Japan}

\date{\today}

\begin{abstract}
We construct the first dynamically stable ergostars (equilibrium neutron
stars that contain an ergoregion) for a compressible, causal equation of
state. We demonstrate their stability by evolving 
both strict and perturbed equilibrium configurations
in full general
relativity for over a hundred dynamical timescales ($\gtrsim 30$ rotational
periods) and observing their stationary behavior. This stability is in 
contrast to earlier models which prove radially unstable to collapse.
Our solutions are highly differentially rotating hypermassive neutron stars 
with a corresponding spherical compaction of $C=0.3$. Such ergostars
can provide new insights into the geometry of spacetimes around highly
compact, rotating objects and on the equation of state at supranuclear
densities. Ergostars may form as remnants of extreme binary neutron star
mergers and possibly provide another mechanism for powering short gamma-ray
bursts.
\end{abstract}

\maketitle

\textit{Introduction.}\textemdash
Two key characteristics of black holes (BHs) are the \textit{event horizon} and the
\textit{ergoregion}. The former represents the ``surface of no return'', i.e. the 
boundary of the region of
spacetime we cannot communicate with (at least in classical theory), while the
latter is a region where there are no timelike static observers and 
all trajectories (timelike or null) must rotate in the
direction of rotation of the BH (frame-dragging). 
For stationary, rotating spacetimes the existence 
of an event horizon implies the existence of an ergoregion, but the opposite is
not true. Ergoregions are associated to two important astrophysical processes which are
both related to the extraction of energy from a rotating BH:
First, as described by Penrose \cite{Penrose:1969pc}, since the energy of a particle as 
seen by an observer at infinity can be negative inside the ergoregion, energy extraction
is possible through a simple decay. Second is the powering of relativistic
jets through the Blandford-Znajek process \cite{BZeffect}. Although according to the
membrane paradigm \cite{Thorne86}, jet
formation is associated with the BH horizon, Komissarov pointed out 
\cite{Komissarov:2004ms,2005MNRAS.359..801K} that the threading of the ergoregion 
by magnetic field lines and the subsequent twisting of them due to frame dragging is 
all that is necessary for the energy creation of a relativistic jet, while a horizon is
not. Preliminary force-free numerical simulations 
of ergostars using the Cowling approximation confirm this hypothesis \cite{Ruiz:2012te}.

A stationary, asymptotically flat spacetime possesses a timelike Killing vector
that asymptotically corresponds to time translations. This vector inside an
ergoregion tips over and becomes spacelike, making the conserved total energy of 
a freely moving particle there negative with respect to the asymptotic
observer. A nonaxisymmetric perturbation that radiates positive energy at 
infinity will make the negative energy in the ergoregion even more negative in
order for the conservation of energy to be satisfied. This will lead to a
cascading instability that was first discovered by Friedman 
\cite{1978CMaPh..63..243F} and recently was put on a rigorous footing by
Moschidis \cite{2018CMaPh.358..437M}. It belongs to the class of ``rotational
dragging instabilities'' whose most famous member is the so-called
Chandrasekhar-Friedman-Schutz (CFS) instability (induced by gravitational-radiation) 
\cite{1970ApJ...161..561C,1978ApJ...221..937F,1978CMaPh..62..247F} valid for any
rotating star, irrespective of its rotation rate. In this paper we call stars that
contain ergoregions  \textit{ergostars}.

The fact that the ergoregion instability was considered ``secondary'' was not
only due to the scarcity of rotating star models exhibiting such behavior, but equally
importantly, due to its very long \textit{secular} ($\gtrsim$ gravitational radiation) timescale 
\cite{1978RSPSA.364..211C, 1996MNRAS.282..580Y, Brito:2015oca}
(see also \cite{Kokkotas:2002sf}). 
Although the existence of ergoregions in rotating stars has been 
questioned \cite{1978MNRAS.182...69S}, they were found by a 
number of authors since the first work of
Wilson \cite{1972ApJ...176..195W}, who employed a compressible equation of state
(EoS), differential rotation, and an assumed density distribution.
Butterworth and Ipser \cite{1975ApJ...200L.103B} and more recently 
Ansorg, Kleinwachter, and Meinel \cite{Ansorg:2001pe} constructed self-consistent,
rapidly rotating, incompressible stars containing ergoregions 
(see also \cite{Ames:2018xqt, Ames:2016coj} for ergoregions in the self-gravitating Vlasov system). 
A larger parameter space was investigated by Komatsu, Eriguchi, and Hachisu 
\cite{1989MNRAS.239..153K} (KEH) who presented self-consistent solutions with a 
polytropic EoS and differential rotation, reaching all the way up to the most
extreme toroidal configurations ($R_p/R_e=0$, where $R_p,\ R_e$ are the polar and
equatorial radii, respectively).

\setlength{\tabcolsep}{5pt}                                                      
\begin{table*}                                                             
\caption{The equilibrium models. 
The polytropic constant used for the 
$\Gamma=3$ models yields a maximum spherical gravitational mass of $4.066 M_\odot$,
which coincides with the maximum spherical gravitational mass of the ALF2cc EoS. 
Parameter $\hat{A}=A/R_e$, where $R_e$ the equatorial radius, determines the
degree of differential rotation, $R_p/R_e$ is the ratio of polar to equatorial
radius, $M_0$ is the rest mass, $M$ is the ADM mass, $J$ is the ADM angular 
momentum, $T/|W|$ is the ratio of kinetic to gravitational energy, ${\rm P_c}$
is the rotational period of the star that corresponds to its central angular velocity
$\Omega_c$, $\Omega_c/\Omega_s$ is the ratio of the central to the surface angular velocity,
and $t_{\rm dyn}\sim 1/\sqrt{\GR}$ the dynamical timescale.}                       
\label{tab:idmodels}                                                     
\begin{tabular}{ccccccccccccc}                                                    
\hline\hline                                                              
Model & EoS   & ER      & $\hat{A}^{-1}$ & $R_p/R_e$ & $M_0\ [\Ms]$   & $M\ [\Ms]$  & $R_e\ [{\rm km}]$    
  & $J/M^2$   & $T/|W|$  & ${\rm P_c}/M$ & $\Omega_c/\Omega_s$  & $t_{\rm dyn}/M$  \\ \hline\hline
iA0.2-rp0.50  & ALF2cc &  \xmark  & $0.2$         & $0.5000$ & $6.683$         &
$5.360$       & $12.62$& $0.8698$ & $0.2266$ & $27.31$  & $1.328$  & $6.9$  \\ \hline
iA0.2-rp0.47  & ALF2cc &  \xmark  & $0.2$         & $0.4688$ & $6.973$         &
$5.587$       & $12.55$& $0.8929$ & $0.2423$ & $25.21$  & $1.359$  & $6.6$  \\ \hline
iA0.2-rp0.45  & ALF2cc &  \cmark  & $0.2$         & $0.4531$ & $7.130$         &
$5.709$       & $12.49$& $0.9035$ & $0.2501$ & $24.18$  & $1.378$  & $6.5$  \\ \hline
iA0.3-rp0.47  & ALF2cc &  \cmark  & $0.3$         & $0.4688$ & $6.900$         &
$5.514$       & $11.52$& $0.8670$ & $0.2354$ & $20.55$  & $1.753$  & $6.7$  \\ \hline
iA0.4-rp0.47  & ALF2cc &  \cmark  & $0.4$         & $0.4688$ & $6.679$         &
$5.334$       & $11.04$& $0.8323$ & $0.2205$ & $17.52$  & $2.216$  & $6.9$  \\ \hline   
g3-iA0.4-rp0.44 & $\Gamma=3$ &  \xmark  & $0.4$         & $0.4375$ & $6.832$   &
$5.761$       & $14.62$& $0.8617$ & $0.2302$ & $20.24$  & $2.027$  & $6.3$  \\ \hline  
g3-iA0.4-rp0.42 & $\Gamma=3$ &  \cmark  & $0.4$         & $0.4219$ & $6.929$   &
$5.845$       & $14.41$& $0.8704$ & $0.2372$ & $19.21$  & $2.073$  & $6.2$  \\ \hline  
g3-iA0.5-rp0.36 & $\Gamma=3$ &  \cmark  & $0.5$         & $0.3594$ & $6.688$   &
$5.718$       & $12.27$& $0.8640$ & $0.2473$ & $13.11$  & $2.876$  & $6.4$  \\ \hline  
\end{tabular}                                                               
\end{table*}                                                                   

The question we want to answer in this Letter is threefold: First, whether any
of the known ergostars with a compressible and causal EoSs are \textit{dynamically} 
stable? If not,
whether the instability is caused by the ergoregion or is it intrinsic to the
other properties of the star.  This is investigated by evolving ergostars
together with nearby equilibria that do not exhibit ergoregions. The whole
analysis is performed in full general relativity and without any approximation,
such as the slow-rotation approximation typically used in perturbation
analysis.  Finally, is it possible to identify any dynamically stable
ergostars?  We will show that all of the models presented in
\cite{1989MNRAS.239..153K} that we have evolved are dynamically unstable and
argue that it will be very difficult, if not impossible, to have stable
ergostars with a simple polytropic EoS. However, we were able to construct a
compressible EoS that leads to dynamically stable ergostars that persist for
our entire integration timescale, which is at least $\sim 20$ ms ($\gtrsim 100$
dynamical times). We present a full general relativistic analysis of multiple
models with this property. 

\textit{Initial data.}\textemdash
Our initial data are constructed with the Cook-Shapiro-Teukolsky (CST) code
\cite{1992ApJ...398..203C} using two equations of state (EoSs). The first one is a $\Gamma=3$
polytrope, which is known to produce differentially rotating ergostars
\cite{1989MNRAS.239..153K}. 
Our motivation was to find stable 
configurations that ideally can represent neutron star (NS) mergers, thus we have chosen to investigate 
the $\Gamma=3$ case since it produced ergostars at higher 
$R_p/R_e$, i.e. with almost spheroidal geometries. A second criterion for our choice
is to find ergostar models with a low  $T/|W|$  so that they are less 
susceptible to nonaxisymmetric instabilities. 
Here $T,\ W$ are the rotational and gravitational potential energy of the stars, respectively.
The second EoS we use is based on the ALF2 EoS \cite{Alford2005}
and denoted as ALF2cc. We replace the region where the rest-mass density 
$\GR_0\geq\GR_{0s}=\GR_{0\rm nuc}=2.7\times 10^{14}\ {\rm gr/cm^3}$ by 
\be
P=\GS(\GR-\GR_s) + P_s \, .
\label{eq:eoscc}
\ee
Here $\GS$ is a dimensionless parameter, $\GR$ is the total energy density, 
and $P_s$ the pressure at $\GR_s$. 
The solutions presented in this work assume $\GS=1.0$, 
i.e. a causal core, which represents the maximally compact, 
compressible EoS \cite{2016PhR...621..127L}.
Apart from a small crust ($\sim 6\% R_e$), the density profiles of all our 
models resemble the ones found in quark stars which exhibit a finite surface
density. In this way we conjecture that it would be possible to construct dynamically
stable quark stars having an ergoregion.
A parameter study for other values of $\GS$, as well as different matching
densities, will be presented elsewhere \cite{Tsokaros_ergostars}.

The differential rotation law is a choice needed 
to solve for hydrostatic equilibrium. We employ the so called ``j-const.'' law 
\cite{1985A&A...146..260E}, which is
written as $j(\Omega)=A^2 (\Omega_c - \Omega)$, where $j$ is the
relativistic specific angular momentum, $A$ is a constant that determines the degree of 
differential rotation and has units of length, and $\Omega_c$ is the angular velocity at the 
center of the star. Other choices like the ones presented in Refs.
\cite{Uryu:2016dqr,Uryu:2017obi} are also possible \cite{Tsokaros_ergostars}.
All our initial models are shown in Table \ref{tab:idmodels}.

\begin{figure*}                                                                   
\begin{center}                                                                   
\includegraphics[width=0.75\columnwidth]{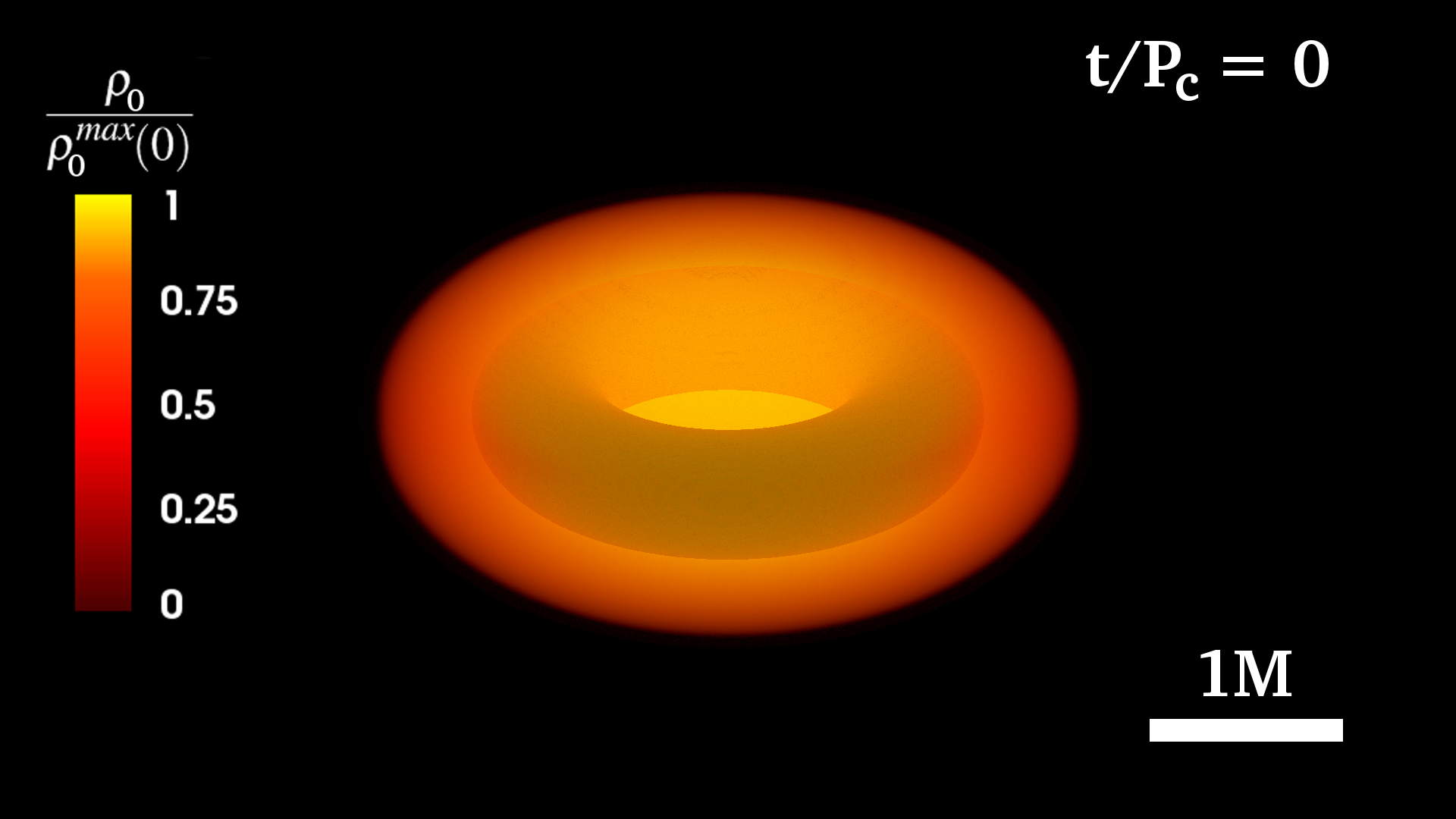}
\includegraphics[width=0.75\columnwidth]{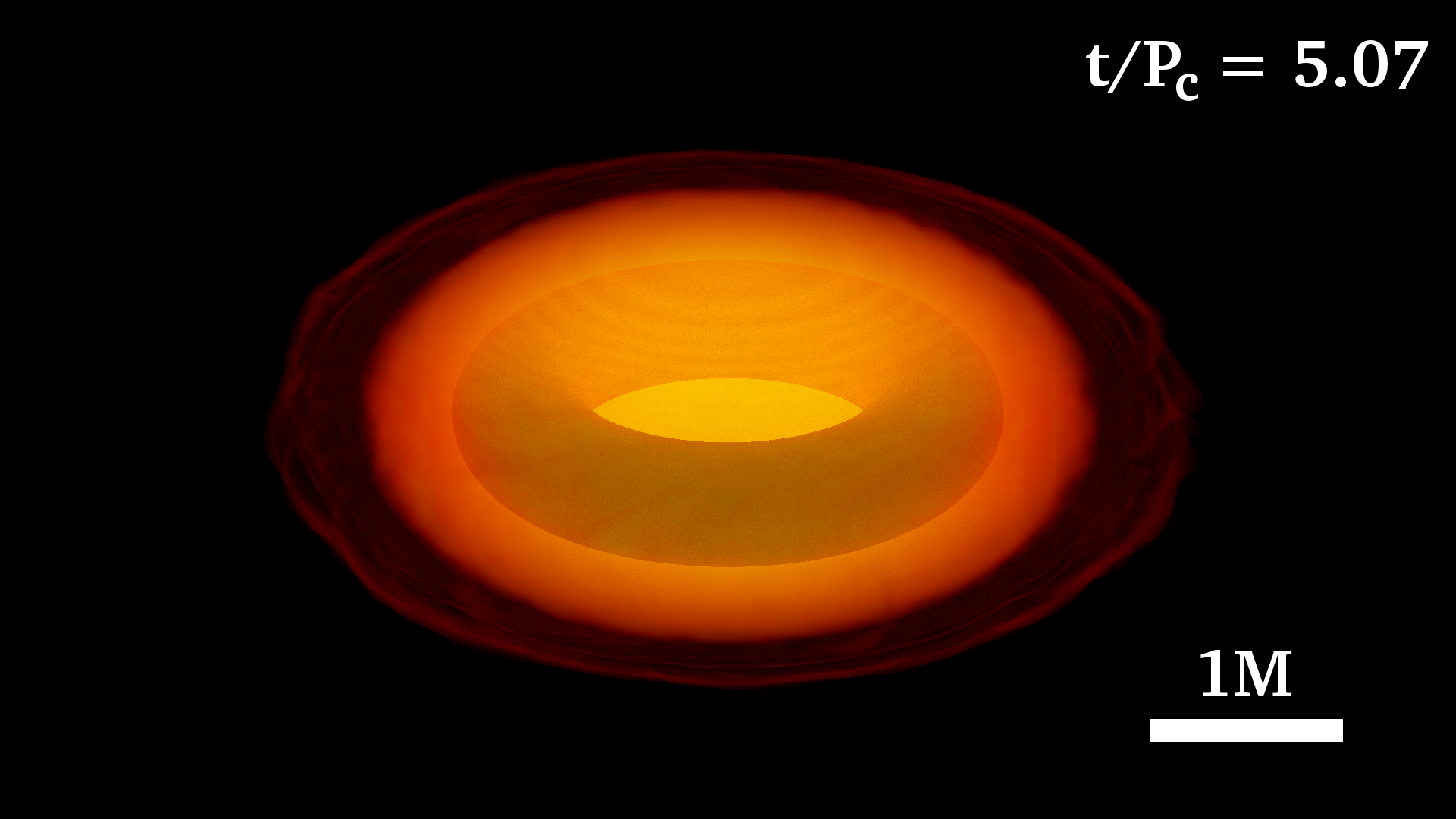}                            
\includegraphics[width=0.75\columnwidth]{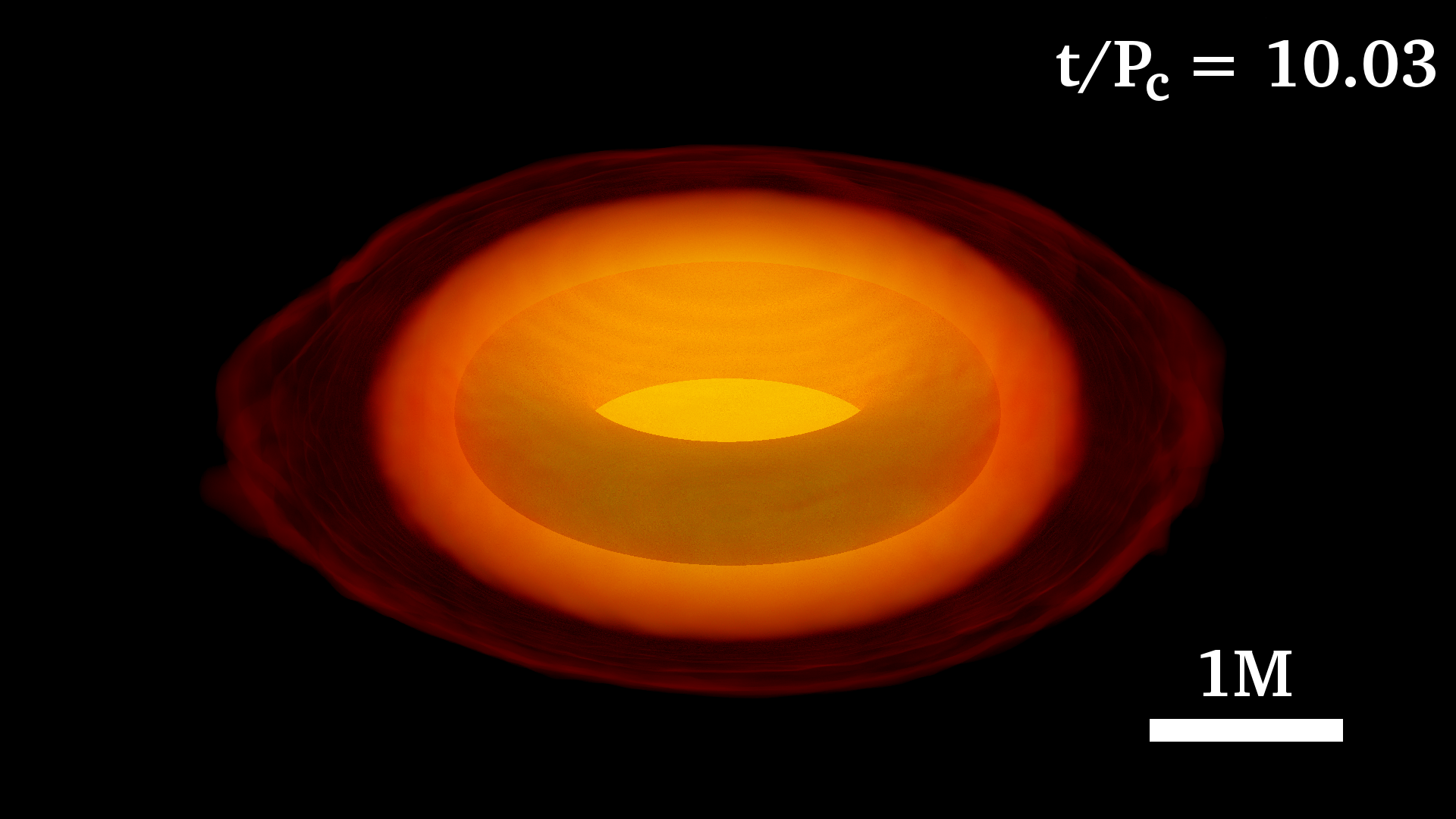}                            
\includegraphics[width=0.75\columnwidth]{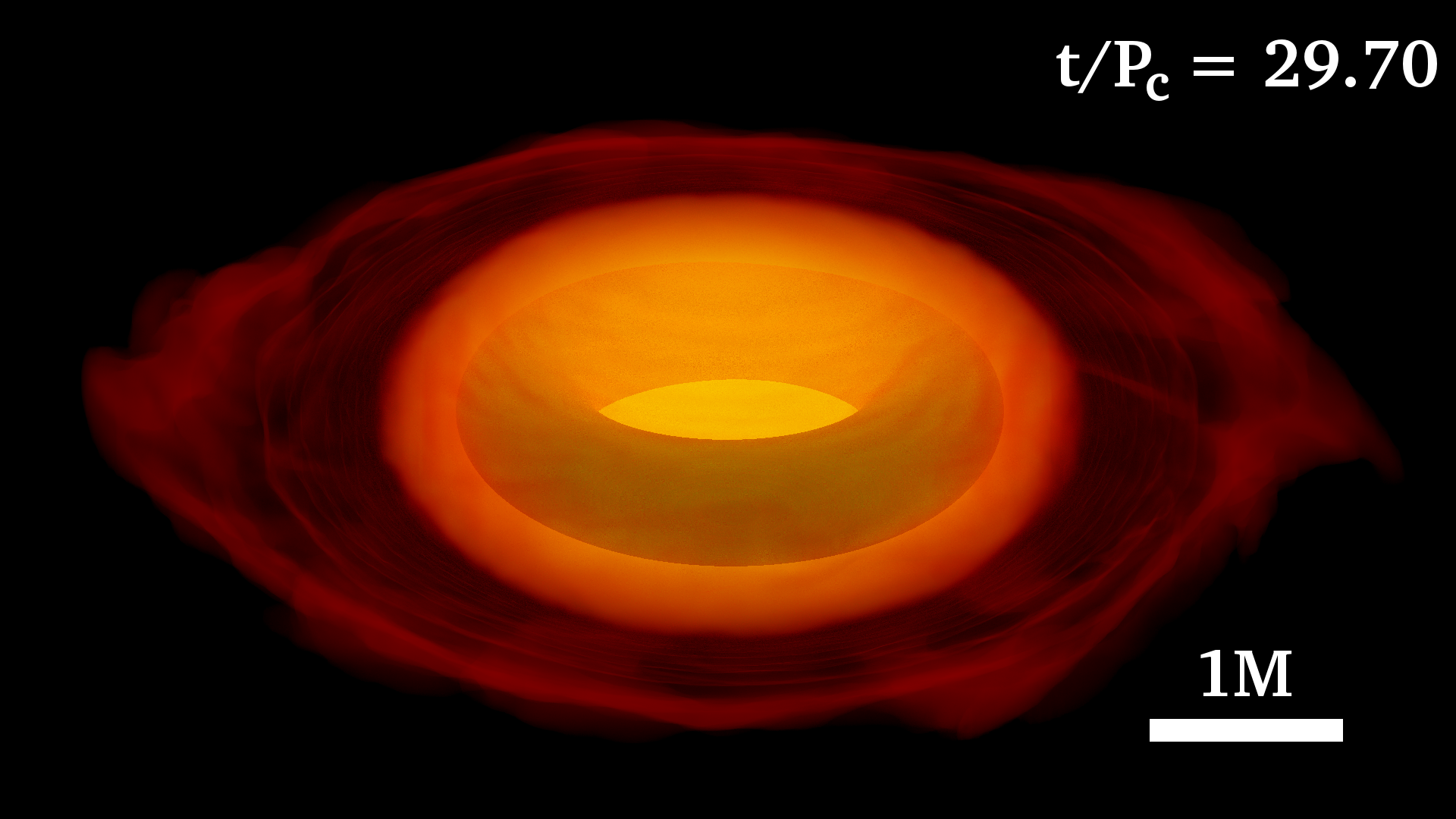}                            
\caption{
Rest-mass density and the ergosurface for the ALF2cc EoS, model iA0.2-rp0.45, 
at 4 different instances of time. The green donut indicates the ergoregion.
Stability is maintained for this equilibrium ergostar.}  
\label{fig:den_ergo}                                                                 
\end{center}                                                                     
\end{figure*}

\begin{figure*}                                                                   
\begin{center}                                                                   
\includegraphics[width=0.75\columnwidth]{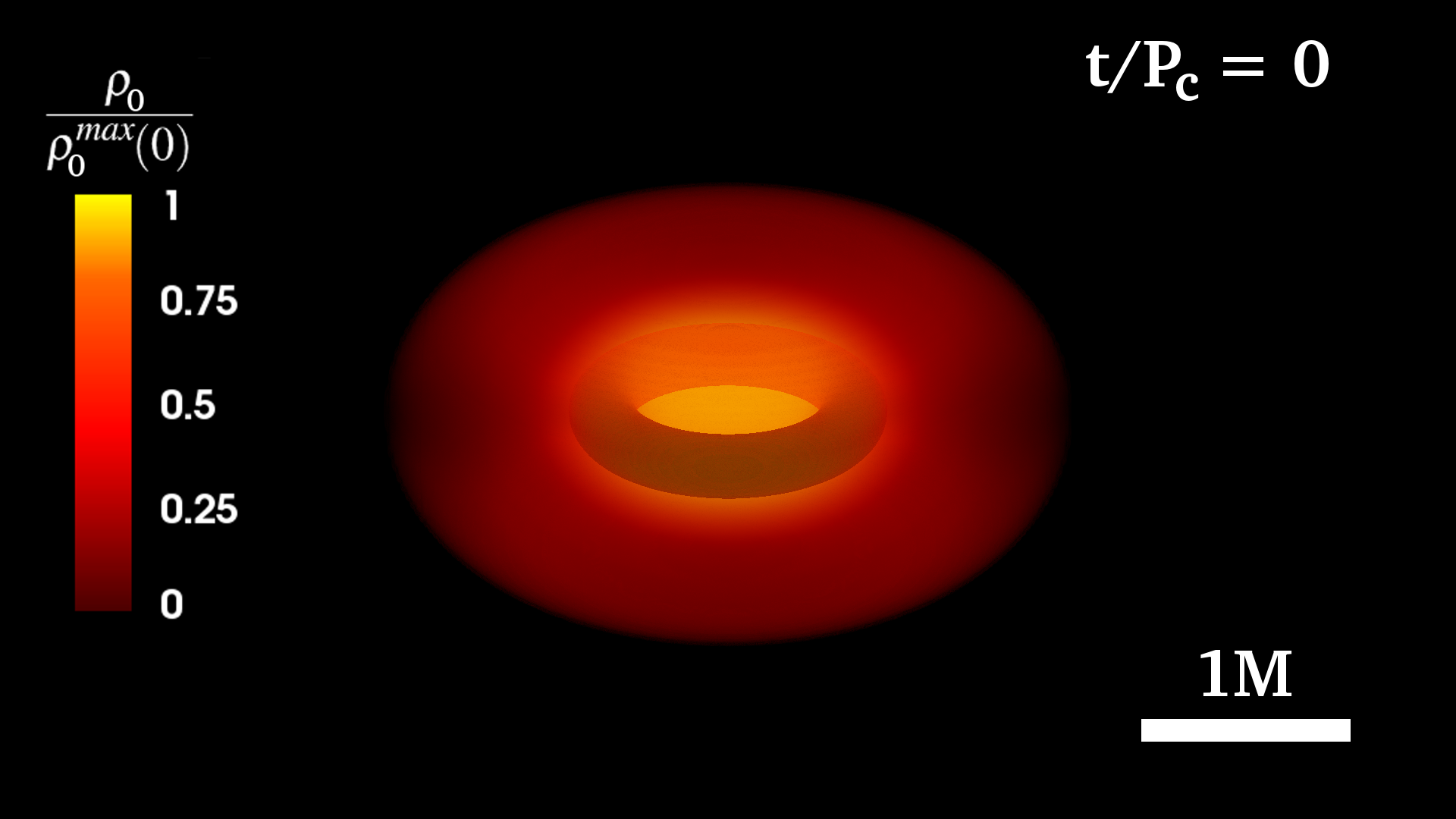}
\includegraphics[width=0.75\columnwidth]{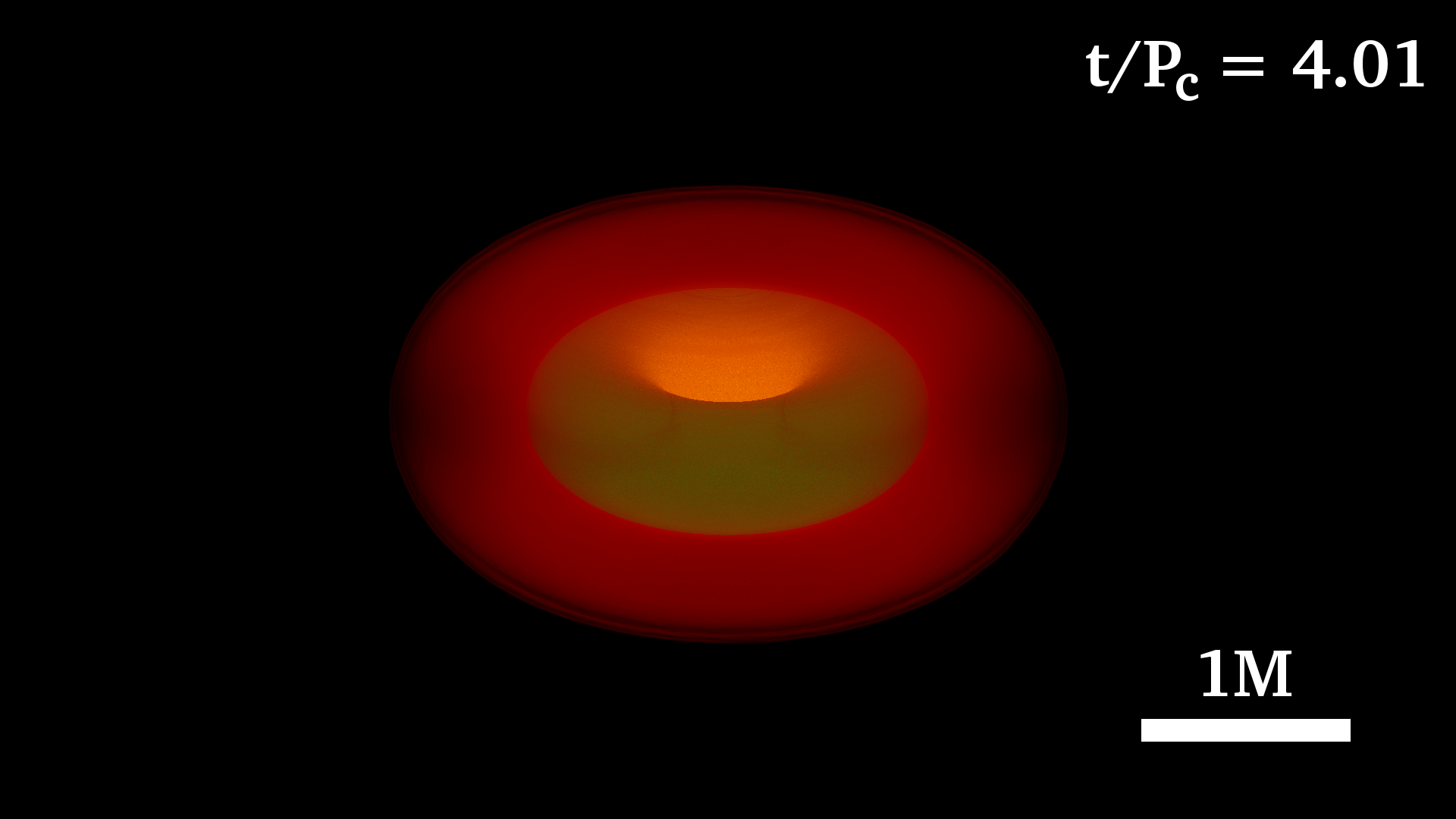}                            
\includegraphics[width=0.75\columnwidth]{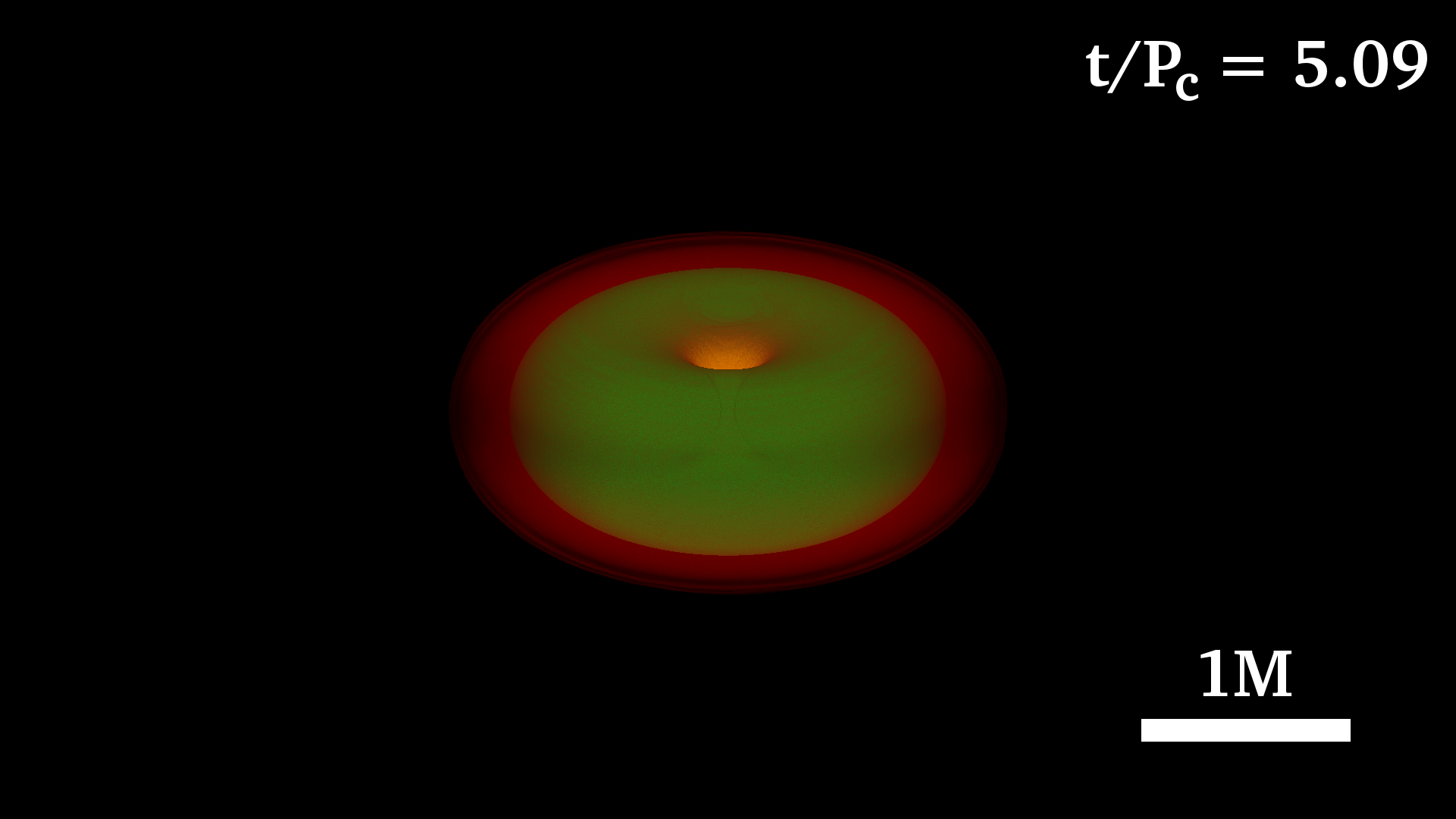}                            
\includegraphics[width=0.75\columnwidth]{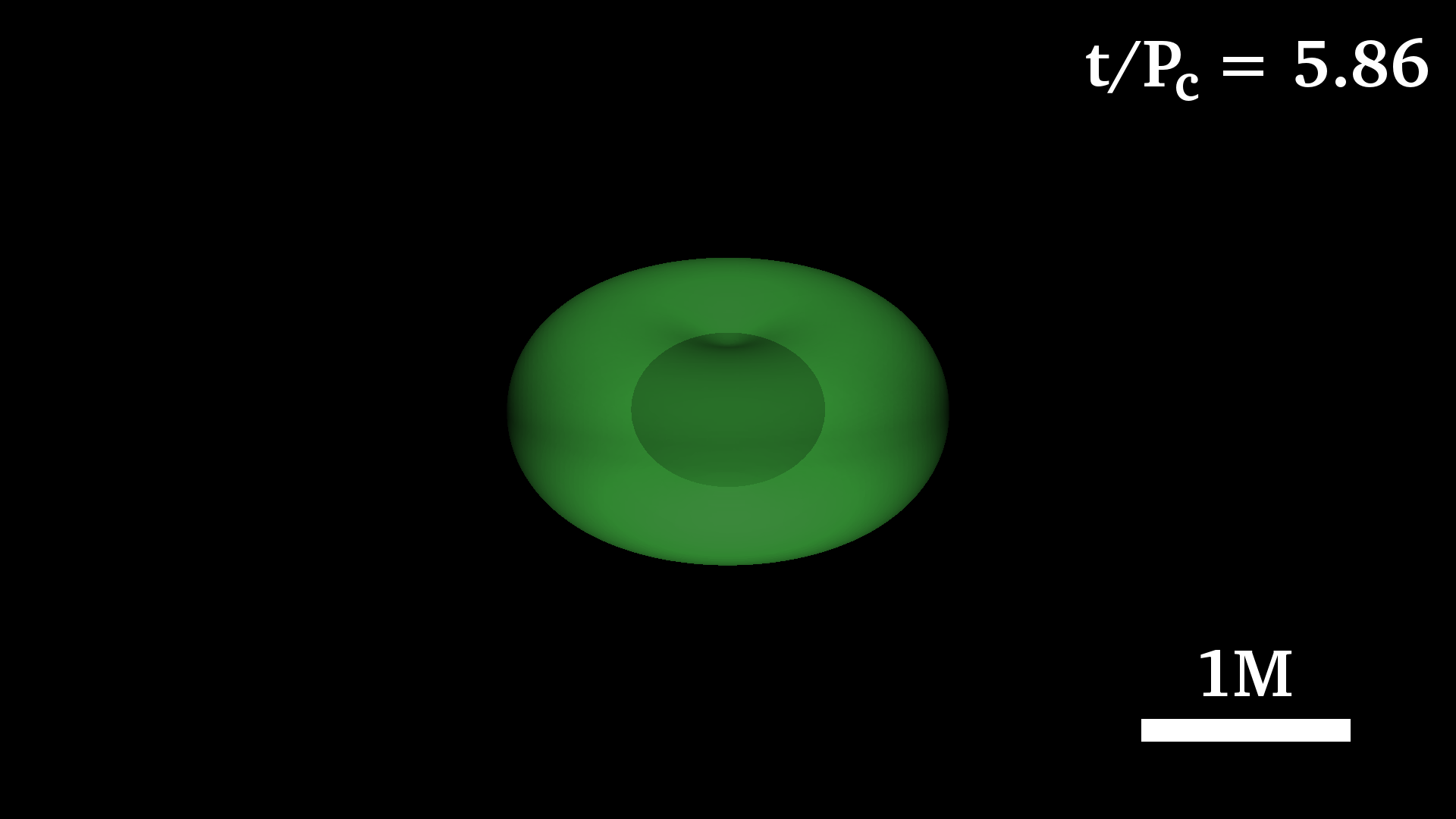}                            
\caption{Similar to Fig. \ref{fig:den_ergo} but for the $\Gamma=3$ EoS model 
g3-iA0.4-rp0.42. This equilibrium ergostar undergoes dynamical collapse to a BH.
The black inner spheroid in the last frame shows the apparent horizon. }  
\label{fig:den_ergo_g3}                                                                 
\end{center}                                                                     
\end{figure*}

\textit{Evolutions.}\textemdash
We use the \illinois adaptive-mesh-refinement code (see e.g.~\cite{Etienne:2010ui}), 
which employs the Baumgarte--Shapiro--Shibata--Nakamura (BSSN) formulation 
of the Einstein's equations~\cite{shibnak95,BS} to evolve the 
spacetime with the standard puncture gauge conditions.
The equations of hydrodynamics are solved in conservation-law form adopting
high-resolution shock-capturing methods. The pressure is decomposed as a sum of a 
cold and a thermal part,                                                
$P = P_{\rm cold} + (\Gamma_{\rm th}-1)\GR_0 (\GE-\GE_{\rm cold})$ 
where $P_{\rm cold}, \GE_{\rm cold}$ are the pressure and specific internal energy 
as computed from the initial data EoS. They are calculated using either a polytropic 
pressure-density relation or Eq. (\ref{eq:eoscc}).
For the thermal part we take  $\Gamma_{\rm th}=5/3$.
The growth of nonaxisymmetric modes is monitored by computing
$C_m = \int \GR_0 u^t \sqrt{-g} e^{im\GP} d^3x $ \cite{Paschalidis:2015mla}. 
In our simulations we used two resolutions, for the ALF2cc models 
with $\Delta x_{\rm min}=153,\ 92$ m. For the $\Gamma=3$ models we used
three resolutions with $\Delta x_{\rm min}=200,\ 140,\ 92$ m. Here 
$\Delta x_{\rm min}$ is the step interval at the finest refinement level.
Note that for the same $\Delta x_{\rm min}$ there is more grid coverage across the star
for the $\Gamma=3$ models because $R_e$ is greater.

\begin{figure*}                                                                   
\begin{center}                                                                   
\includegraphics[width=0.67\columnwidth]{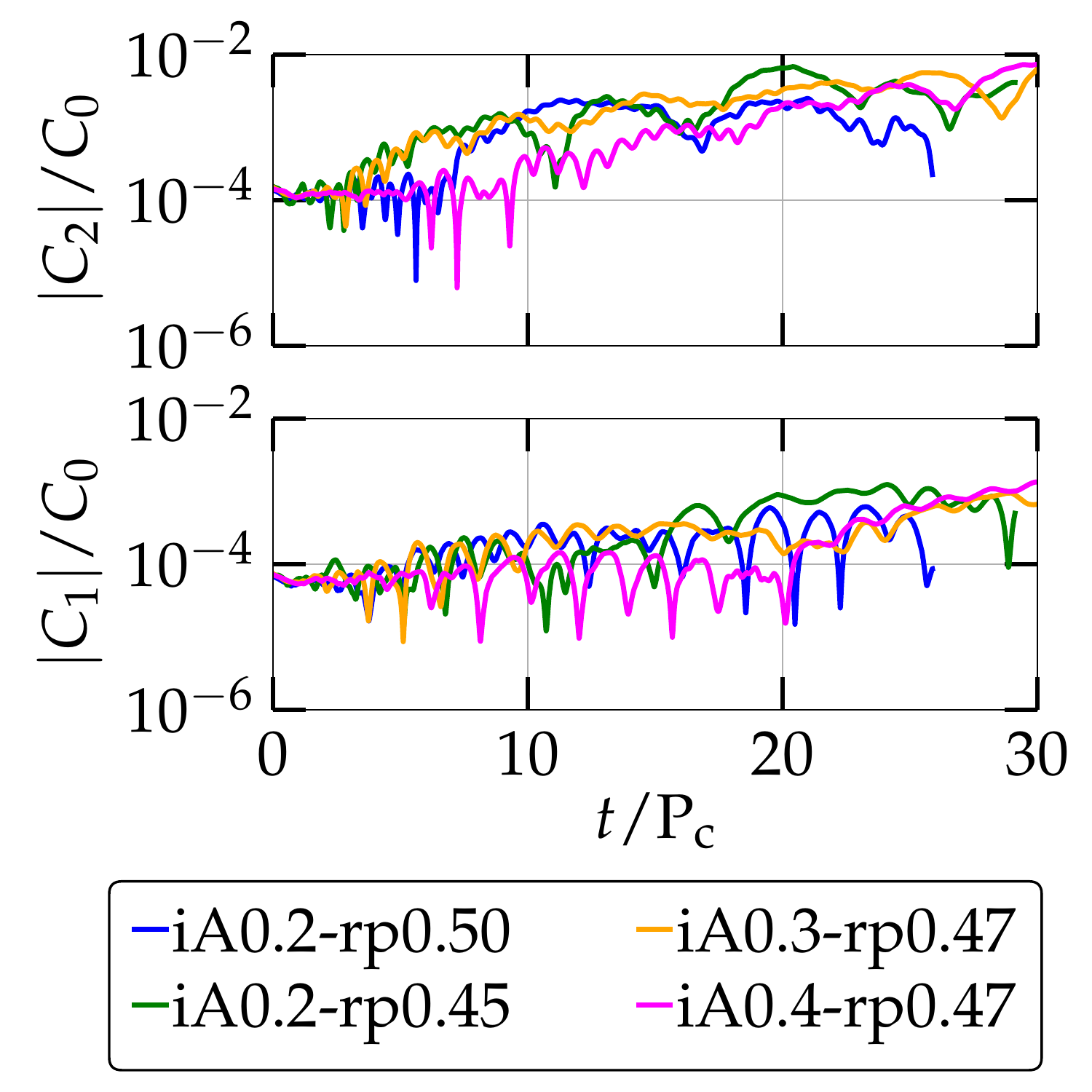}
\includegraphics[width=0.67\columnwidth]{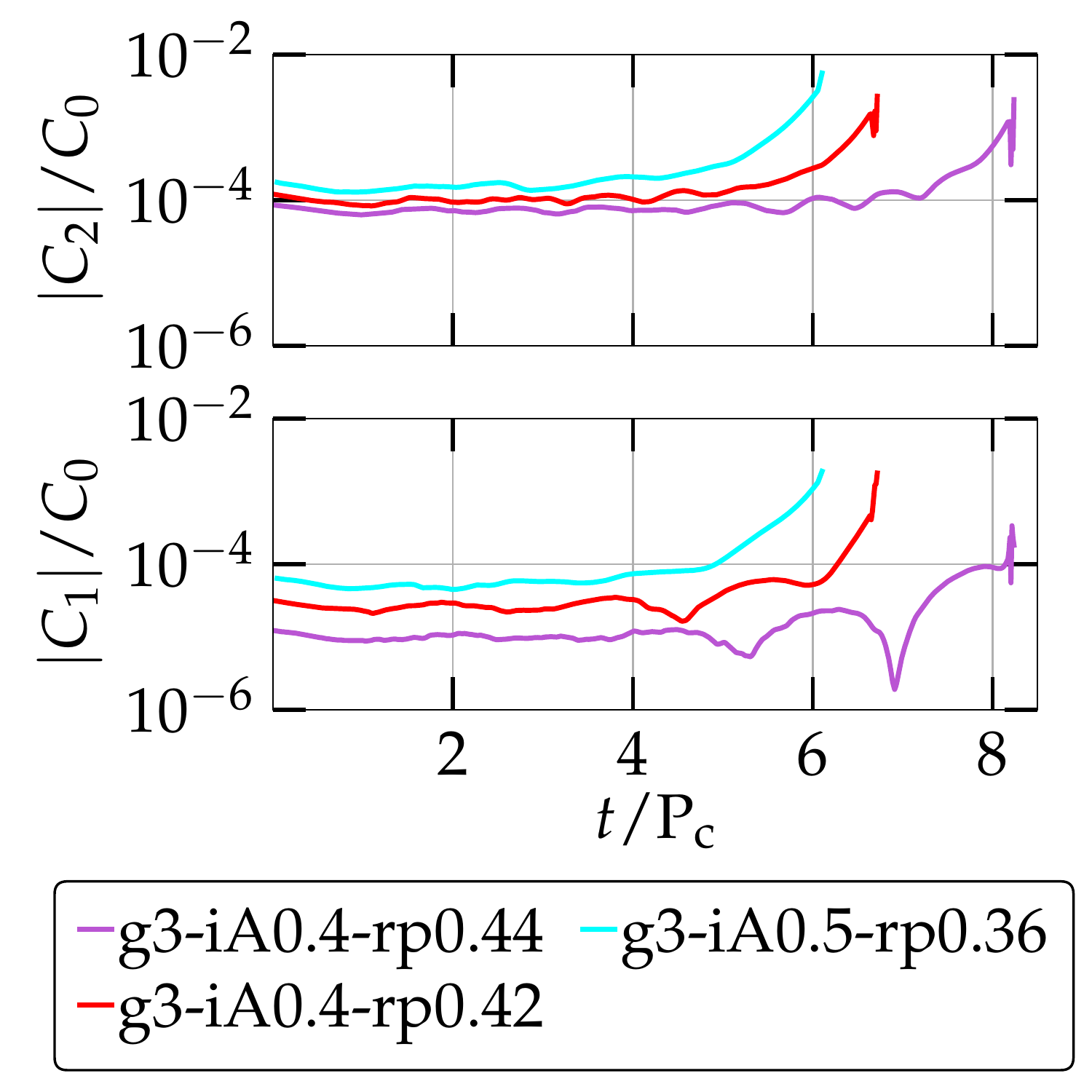}                            
\includegraphics[width=0.67\columnwidth]{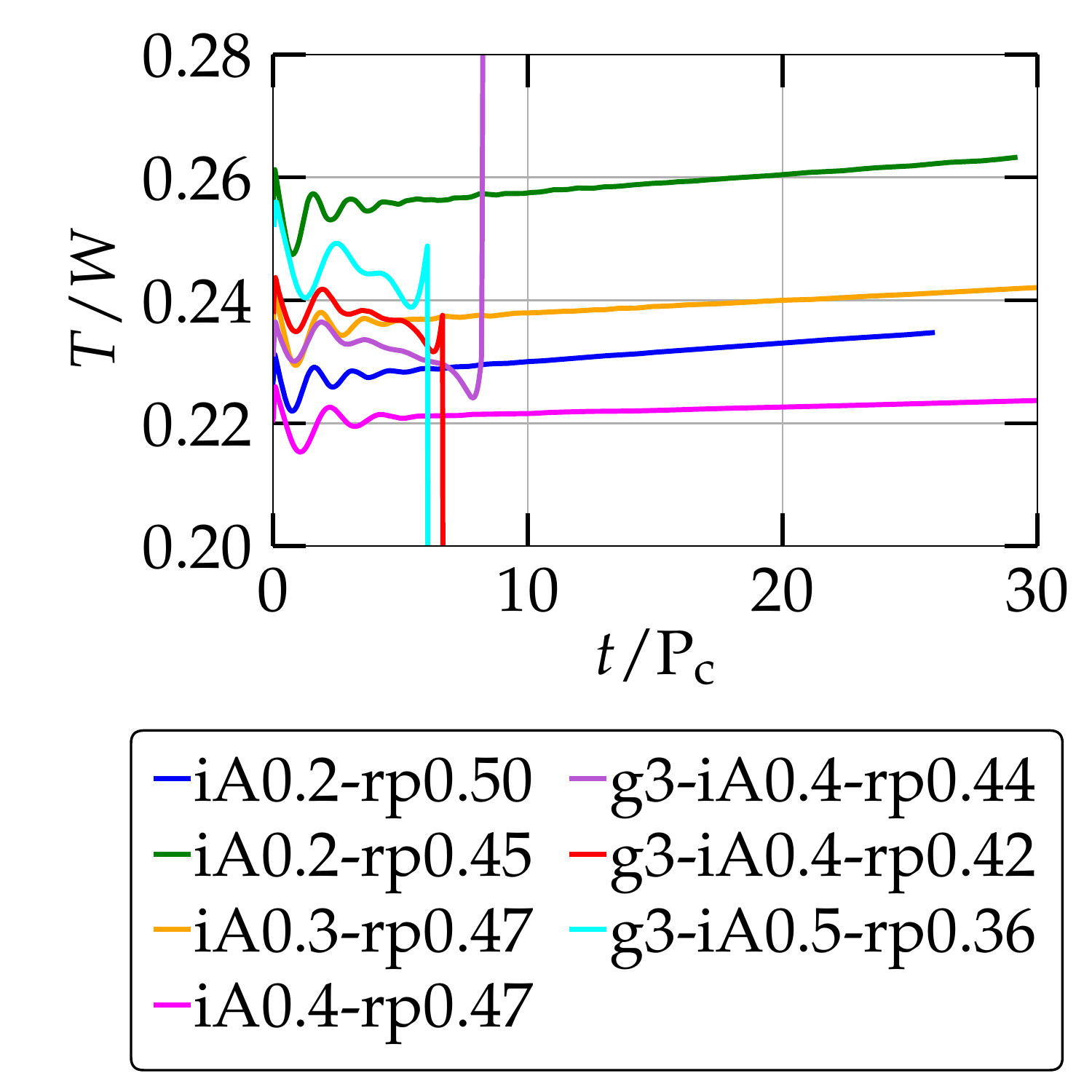}                            
\caption{
Time evolution of the $m=1,2$ modes for the ALF2cc EoS models (left panel), the 
$\Gamma=3$ EoS models (middle panel), and $T/W$ (right panel). The corresponding
dynamical timescales are listed in Table \ref{tab:idmodels}. }  
\label{fig:modes}                                                                 
\end{center}                                                                     
\end{figure*}

Snapshots during the evolution of the ergostars with the ALF2cc and the 
$\Gamma=3$ EoSs are depicted in Figs. \ref{fig:den_ergo}
and Fig. \ref{fig:den_ergo_g3} where two prime examples of each category are
plotted. Fig. \ref{fig:den_ergo} shows the normalized rest-mass density as well as the
ergosurface ($g_{tt}=0$, inner green donut) of the model iA0.2-rp0.45  at 4
instances $t/{\rm P_c}\approx 0,\ 5,\ 10,\ 30$ and constitutes our prime, dynamically stable
ergostar using the ALF2cc EoS that exhibits a causal core, Eq. (\ref{eq:eoscc}). 
As it is clear from that figure the star retains both its axisymmetric structure 
as well as the geometry of the ergoregion for the whole period of our evolution that
reaches approximately 30 rotation periods or 100 dynamical timescales.
This ergostar is the first member that
exhibits an ergoregion along a constant rest-mass (central) density 
$\GR_0=4.52\times 10^{14}\ {\rm gr/cm^3}$ sequence with a decreasing $R_p/R_e$ ratio 
and the j-const law with $\hat{A}=5$.
All equilibrium models before that (i.e. for larger ratios of $R_p/R_e$) do not
contain any ergoregions, while all models after that, i.e. for greater deformations 
(smaller ratios of $R_p/R_e$), contain ergoregions whose size increases with 
increasing deformation. In other words, for the particular sequence of rest-mass 
density and differential rotation law, ergostar iA0.2-rp0.45 is (a) the most 
spheroidal, (b) has the lowest $T/W$, and (c) has the smallest ergoregion.
Note that $T/W=0.25$, which is certainly at the boundary of dynamical stability
\cite{Baumgarte:1999cq, Shibata_2000}. Less deformed models iA0.2-rp0.50 and iA0.2-rp0.47
belong to the same sequence as the ergostar iA0.2-rp0.45 and have the same differential
rotation law but contain no ergoregions. These normal star equilibria have also a smaller
value of $T/W$, and our simulations confirm that they are dynamically stable similarly.

Fig.~\ref{fig:modes} left panel, shows the growth of nonaxisymmetric modes for normal
star iA0.2-rp0.50 as well as ergostars iA0.2-rp0.45, iA0.3-rp0.47,  iA0.4-rp0.47 using
$\Delta x_{\rm min}=153$ m. The same behavior is observed at higher resolution with 
$\Delta x_{\rm min}=92$ m. Evidently the evolution of all stars maintains axisymmetry
on dynamical timescales. 
Particularly
during the last 10 rotation periods both the normal star iA0.2-rp0.50 
and the ergostar iA0.2-rp0.45 (which is shown also in Fig. \ref{fig:den_ergo}) show
a saturation of the $m=1,2$ growth amplitude. Ergostars iA0.3-rp0.47 and iA0.4-rp0.47
have the same central density as iA0.2-rp0.45 but larger differential rotation:
$\hat{A}=3.33$ and $2.5$ respectively.
In the Supplement we present additional evidence for the dynamical stability of these
models by seeding them with an $m=1$ or $m=2$ density perturbation and inspecting their non-growth
in the timescale of our simulations. In addition we show that these stars are stable to 
quasiradial density perturbations.

Fig.~\ref{fig:den_ergo_g3} shows the normalized rest-mass density and ergosurface 
for the $\Gamma=3$ EoS ergostar g3-iA0.4-rp0.42 evolved using $\Delta x_{\rm min}=200$ m
at 4 instances $t/{\rm P_c}\approx 0,\ 4,\ 5$, and at BH formation. Although the 
criterion  $\mathbf{t}\cdot\mathbf{t}=g_{tt}=0$ (where $\mathbf{t}=\partial_t$ is the time 
coordinate basis vector) for ergoregion identification does not strictly hold in the 
nonstationary spacetime of the collapsing star, it is still a reasonable measure given the 
stationary initial and final gravitational equilibria. This model is the first member that
exhibits an ergoregion along a  constant rest-mass density $\GR_0=3.846\times 10^{14}\ {\rm gr/cm^3}$ 
sequence with $\hat{A}=2.5$. All equilibrium models with less deformation do not contain any 
ergoregions, while all models with larger deformations contain larger size ergoregions. 
Also ergostar g3-iA0.4-rp0.42 \textit{is less deformed and has smaller $T/W$} than any of the
$\Gamma=3$ models of Ref. \cite{1989MNRAS.239..153K}, therefore is less prone to bar-mode
instabilities. Other models in Ref. \cite{1989MNRAS.239..153K} containing ergoregions have 
very small ratios of $R_p/R_e$ and much higher $T/W$, thus the possibility of being
dynamically unstable as well is much higher. This was indeed proven recently in a select number of such
extreme toroids in \cite{Espino:2019xcl}.
Fig. \ref{fig:modes} middle panel shows the growth of nonaxisymmetric modes for the $\Gamma=3$ 
EoS models g3-iA0.4-rp0.44 (normal star), g3-iA0.4-rp0.42 (ergostar shown in 
Fig. \ref{fig:den_ergo_g3}) and g3-iA0.5-rp0.36 (also an ergostar) until just after BH 
formation. The small values of $C_m/C_0$ imply the free-fall 
collapse of those models is axisymmetric. The resolution used is $\Delta x_{\rm min}=140$ m. 
In the right panel of Fig. \ref{fig:modes} we plot $T/W$ for all the models discussed above.
As it is evident the $\Gamma=3$ models all collapse while $T/W$ slightly decreases from
their initial values. Also ergostar iA0.2-rp0.45 has the largest $T/W$ in the ALF2cc EoS 
set of models while the ergostar with the highest degree of differential rotation, iA0.4-rp0.47, 
has the smallest. The radial instability of the $\Gamma=3$ EoS models of Table \ref{tab:idmodels} 
is verified by using three different resolutions with the highest one having $\Delta x_{\rm min}=92$ m. 
The evolution of the shape of the ergosphere for the model g3-iA0.4-rp0.42 is presented in the Supplement.

\textit{Discussion.}\textemdash
In this Letter we presented dynamically stable equilibrium rotating NSs that contain ergoregions. 
The EoS that we employed is causal at the core and ALF2 at the outer layers of the star. We also 
proved that previously calculated polytropic ergostars are dynamically unstable. The secular 
evolution of our models will probably be determined by the Friedman instability \cite{1978CMaPh..63..243F} 
in the absence of other dissipative mechanisms. Despite that, and given the long timescales involved, 
the possibility of existence of such equilibria raises a number of questions, the most obvious of 
them being the fate of ergostars exhibiting internal dissipative mechanisms, such as viscosity or 
magnetic fields (which may serve as turbulent viscosity). Preliminary calculations of magnetic 
effects in fixed spacetimes \cite{Ruiz:2012te} have shown that such systems can launch jets similar
to BHs surrounded by magnetized disks. If the merger of two NSs forms an ergostar remnant
which can launch a jet, the timescale for jet formation will be earlier than the one for a normal 
hypermassive NS \cite{Ruiz:2016rai,Ruiz:2018wah}. This feature may have consequences in the 
theoretical analysis of events like GW170817 and its short gamma-ray burst counterpart GRB 170817A. 
Such open problems, as well as questions related to the range of EoSs and differential rotating laws 
that can lead to ergostars, or the possibility of binary ergostar remnants, are under investigation.

Movies highlighting results of our simulations can be found at
\texttt{http://research.physics.illinois.edu/} \texttt{cta/movies/Ergostar/}.

\textit{Acknowledgments.}\textemdash
It is a pleasure to thank R. Haas and V. Paschalidis for useful discussions. We 
also thank the Illinois Relativity group REU team, G. Liu, K. Nelli, and  M. N.T
Nguyen for assistance in creating Figs.~\ref{fig:den_ergo} and~\ref{fig:den_ergo_g3}.
This work was supported by NSF grant PHY-1662211 and NASA grant  80NSSC17K0070 to the
University of Illinois at Urbana-Champaign, as well as by JSPS Grant-in-Aid for Scientific
Research (C) 15K05085 and 18K03624 to the University of Ryukyus. This work made use of the
Extreme Science and Engineering Discovery Environment (XSEDE), which is supported by National
Science Foundation grant number TG-MCA99S008. This research is part of the Blue Waters
sustained-petascale computing project, which is supported by the National Science Foundation
(awards OCI-0725070 and ACI-1238993) and the State of Illinois. Blue Waters
is a joint effort of the University of Illinois at Urbana-Champaign and its National Center
for Supercomputing Applications.  Re-sources supporting this work were also provided by the
NASA High-End Computing (HEC) Program through the NASA Advanced  Supercomputing  (NAS)
Division at Ames Research Center.   

\bibliographystyle{apsrev4-1}
\bibliography{references}

\section{Supplemental Material}

\section{Numerical stability analysis}

\begin{figure}
\begin{center}
\includegraphics[width=0.98\columnwidth]{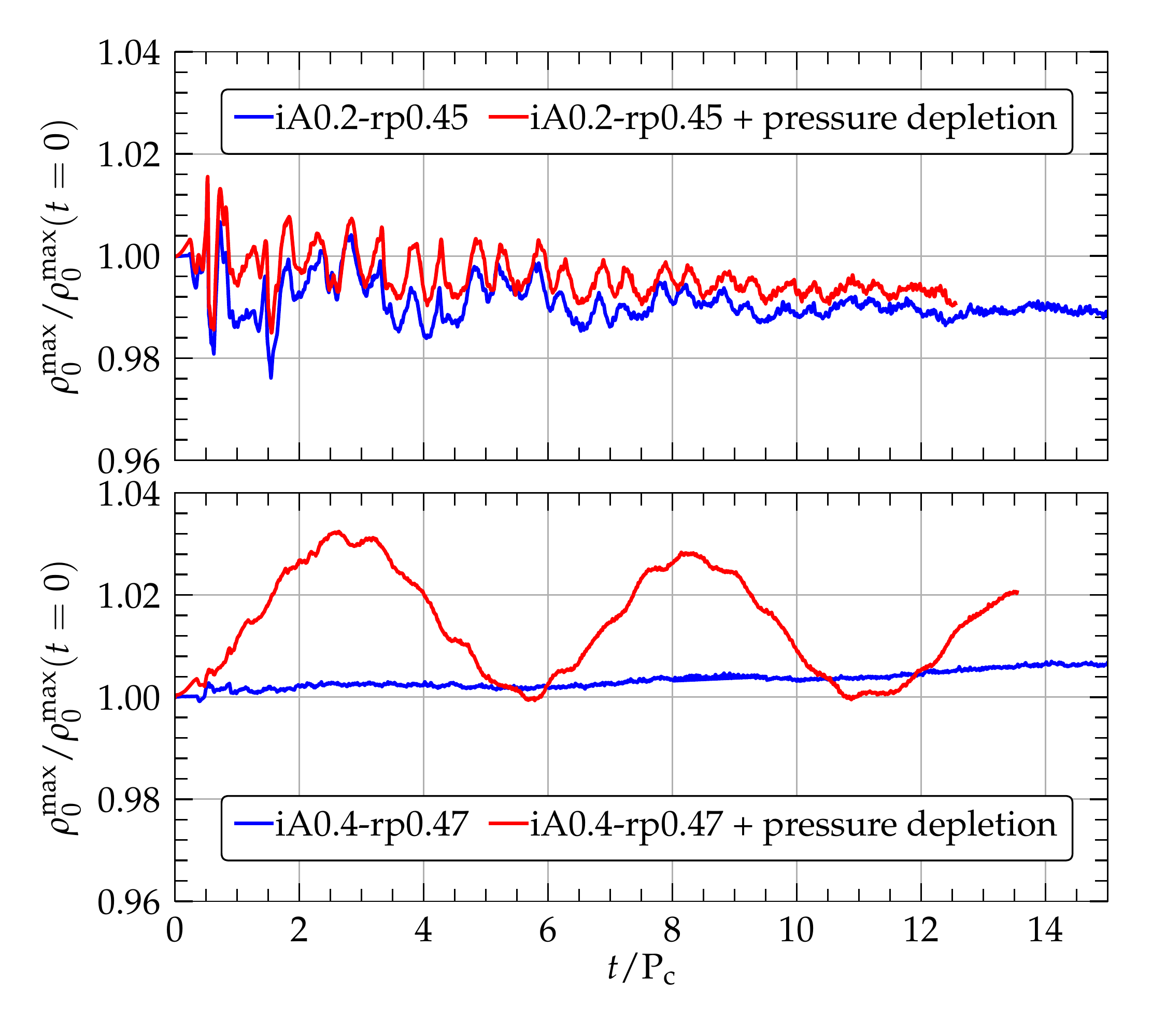}
\caption{
Top panel: model iA0.2-rp0.45. Maximum density evolution for the equilibrium 
configuration as well as the one with pressure depletion. Bottom panel: similarly for 
model iA0.4-rp0.47.}
\label{fig:ergo_pert_pdep}
\end{center}
\end{figure}

\begin{figure}
\begin{center}
\includegraphics[width=0.98\columnwidth]{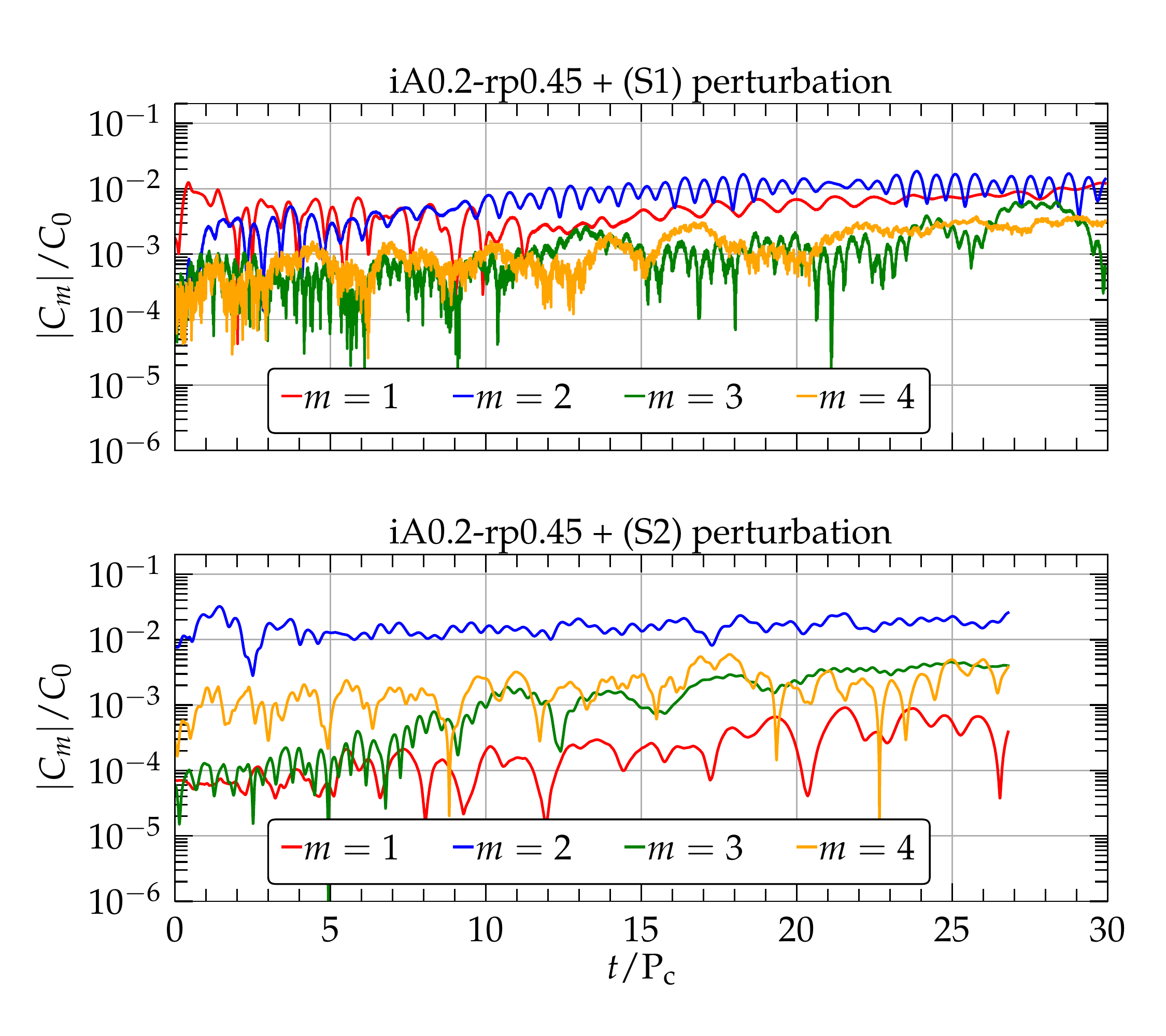}
\caption{
Top panel: mode growth for the model iA0.2-rp0.45 seeded with an 
$m=1$ perturbation. Bottom panel: similarly for an $m=2$ perturbation.}
\label{fig:ergo_pert}
\end{center}
\end{figure}

In this section we further probe the dynamical stability of our ergostar models 
iA0.2-rp0.45, iA0.3-rp0.47, iA0.4-rp0.47 by exciting a number of density
perturbations in them. By applying an $m=0$ perturbation we investigate
the stability against quasiradial oscillations, while an $m=1$ or $m=2$
perturbation tests the stability against nonaxisymmetric modes. 
The $m=0$ case is
implemented by depleting the pressure in the stars by a certain 
amount which we chose to be $1\%$. The $m=1$ perturbation
is implemented by modifying the density profile according to 
\be
\rho_0 \rightarrow \rho_0\left(1+B\frac{x+y}{R_{\rm eq}}\right)\ ,  \tag{S1}
\label{eq:pert1}
\ee
where $R_e$ is the equatorial radius of the star, and $B$ a constant
that we take to be $5\%$. Finally, the $m=2$
perturbation is applied by the use of the transformation
\be
\rho_0 \rightarrow \rho_0\left(1+B\frac{x^2-y^2}{R_{\rm eq}^2}\right) \ , \tag{S2}
\label{eq:pert2}
\ee
for the same choice of $B$.
In order to isolate the effects that are coming from the ergoregion we 
apply the same perturbations to the regular stars iA0.2-rp0.50, iA0.2-rp0.47
that do not contain an ergoregion.
Overall all our models behave in the same \textit{stable} way and we could 
not identify any peculiar behavior that could in principle be attributed to
the existence of the ergoregion.

In Fig. \ref{fig:ergo_pert_pdep} we show the maximum density evolution for the 
equilibrium as well as the pressure-depletted stars 
iA0.2-rp0.45 (top panel) and  iA0.4-rp0.47 (bottom panel). The equilibrium 
models do not exhibit any significant oscillations, therefore the pressure-depleted 
ones are stable as well. The featured ergostar, Fig. 1, whose
maximum density coincides with its geometric center, exhibits very small
oscillations when pressure depleted. On the other hand star iA0.4-rp0.47 
whose maximum density is off-center oscillates more. Also, the slight increase
in the density for the equilibrium model is due to numerical viscosity,
as proved by evolving with different resolutions.
Overall, all models in Table I present the same behavior when we pressure-deplete 
them, therefore they are all \textit{stable} against quasiradial perturbations
on dynamical timescales.

In Fig. \ref{fig:ergo_pert} we present the effects of the $m=1$ one-arm perturbation
Eq. (\ref{eq:pert1}) in the top panel, and the effects of the $m=2$ bar-mode perturbation
Eq. (\ref{eq:pert2}) in the bottom panel, for the featured ergostar iA0.2-rp0.45. 
Plotted is the evolution of the first four modes. Model iA0.2-rp0.45
has the largest $T/W$ and therefore is more prone to the bar-mode instability.
In addition as we can see from Fig. 3, right panel, this ergostar has 
$T/W(t> 20 {\rm P_c})>0.26$ which in turn suggests that when a bar mode is excited 
the possibility of exponential growth on a dynamical timescale is significant. The bottom panel
of Fig. \ref{fig:ergo_pert} shows that this intuition is mistaken. The $m=2$
perturbation shows no sign of growth whatsoever for the time of our integrations.
The mode that mostly grows is the $m=3$ but still it has a small amplitude.
The excitation of an $m=1$ mode on the other hand instigates the development 
of an $m=2$ mode as well. Both modes grow at the level of $1\%$ by the end
of our simulations and, as seen in Fig. \ref{fig:ergo_pert},
they also show no sign of exponential growth. 
Almost identical behavior is observed in all models of Table I with the 
ALF2cc EoS. The facts that we do not observe any essential growth of any modes, 
as well as there is no geometrical or topological change of the equilibrium
models for \textit{many dynamical times} leads us to conclude that all of our ALF2cc 
stars are \textit{dynamically stable}.

\section{Ergosphere evolution}

\begin{figure}
\begin{center}
\includegraphics[width=0.98\columnwidth]{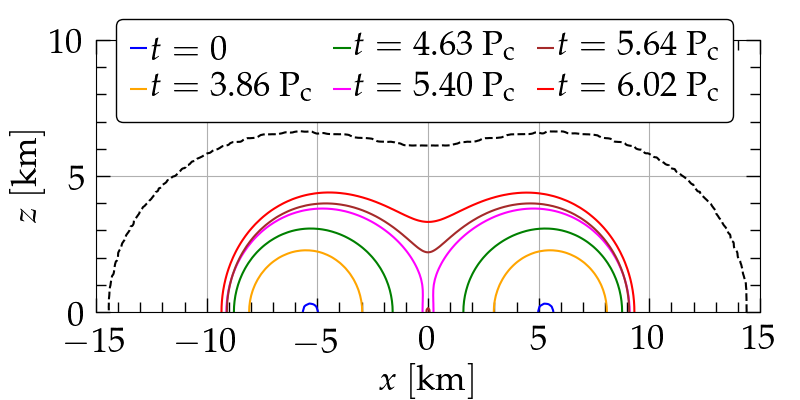}
\caption{
Ergoregion contours at different times for collapsing ergostar g3-iA0.4-rp0.42.
The blue contour ($t=0$) shows the ergoregion for the initial data
while the red contour ($t=6.02\ {\rm P_c}$) corresponds to the
ergoregion of the final stationary BH. Black dashed line depicts initial star surface.}
\label{fig:ergo_evol}
\end{center}
\end{figure}

Fig. \ref{fig:ergo_evol} shows the ergosurface at 6 different instances of time for the
collapsing ergostar g3-iA0.4-rp0.42 shown in Fig. 2. The ergoregion smoothly transitions from a
toroidal to a spheroidal topology
around $t \approx 5.6\ {\rm P_c}$, while the apparent horizon appears at 
$t=5.83\ {\rm P_c}$. The final BH has $a:=J/M^2=0.87$ which coincides with the corresponding
value of the ergostar in Table I. 

%
%

\end{document}